\documentclass[referee]{raa}            

\usepackage{graphicx,times}             
\usepackage{natbib}
\usepackage{amssymb,amsmath}
\bibpunct{(}{)}{;}{a}{}{,}

\usepackage[pagebackref=true]{hyperref}

\begin{document}

  \title{Mass distribution of neutron stars in binary systems
}

   \volnopage{Vol.0 (20xx) No.0, 000--000}      
   \setcounter{page}{1}          

   \author{Zhe Hu 
      \inst{1}
   \and GuoLiang Lü
      \inst{1,2}
   \and ChunHua Zhu
      \inst{1}   
   \and Sufen Guo
      \inst{1}
   \and Helei Liu
      \inst{1}
    \and Lin Li
      \inst{1}
    \and Zhuowen Li
      \inst{1}
    \and Zhenwei Li
      \inst{3}
   }

   \institute{School of Physical Science and Technology, Xinjiang University, Urumqi 830046, China; guolianglv@xao.ac.cn, chunhuazhu@sina.cn
        \and
             Xinjiang Astronomical Observatory, Chinese Academy of Sciences, Urumqi 830011, China
Received 2024 October 31; revised 2025 January 15; accepted 2025 January 19; published 2025 February 28
        \and
        Yunnan Observatories, Chinese Academy of Sciences, Kunming, 650011, People’s Republic of China\\
\vs\no
   {\small Received 20xx month day; accepted 20xx month day}}

\abstract{It is known that the mass distribution of the known neutron stars (NSs) exhibits a bimodal pattern. The origin of this distribution remains a subject of debate. We constructed a super-Eddington accretion model for accreting neutron stars and investigated the mass growth and distribution of these stars using the population synthesis method. We find, in our model, the mass growth of NSs depends on the binary orbital period and the mass of the donor star. Our results can successfully account for the bimodal distribution of NS masses. The peak distribution of NS masses at around $\sim 1.8\,M_{\odot}$ primarily originates from NS binary systems where the donor star mass is less than $\sim 1.6\,M_{\odot}$ and the orbital period is shorter than 20 days; while, NS systems that may undergo common envelope evolution and these NSs can account for the mass peak at $1.4\,M_{\odot}$.
}

   \authorrunning{H.Z}            
   \titlerunning{ Mass distribution of neutron stars in binary systems }  

   \maketitle

%
%
\section{Introduction}           
\label{sect:intro}

NSs are one of the final products of stellar evolution, and the study of their masses serves as a powerful tool for investigating supernova physics, the evolution of compact binary systems, and the equation of state of neutron matter. Early observations indicated that NS masses follow a Gaussian distribution, with a mean of approximately \( 1.35 \, M_{\odot} \) and a dispersion of \( 0.04 \, M_{\odot} \) \citep{thorsett1999neutron}. Due to precise mass measurements of the Hulse–Taylor binary pulsar \citep{taylor1979measurements,taylor1992experimental}, a value of \( 1.4 \, M_{\odot} \) has been adopted as the standard in most studies.

It is well known that the NSs whose masses have been measured reside in binary systems, and their mass distribution is described by a bimodal model (see, e.g., \citealt{ozel2012mass, pejcha2012observed, rocha2023mass}).
From a theoretical perspective, explaining the origin of the \( 1.8 \, M_{\odot} \) peak in the NS mass distribution from the birth of NSs remains controversial \citep{muller2016simple, sukhbold2018high, woosley2020birth}. 
Stars with initial masses between approximately \( 8 \) and \( 10 \, M_{\odot} \) evolve, due to mass loss or binary interactions, to form a degenerate oxygen-neon-magnesium core with a mass between \( 1.83 \) and \( 2.25 \, M_{\odot} \) (depending on the metallicity of the star), which triggers electron capture on \( ^{24}\text{Mg} \) and \( ^{20}\text{Ne} \), leading to core collapse and the formation of a NS \citep{nomoto1984evolution, doherty2015super}. 
According to theoretical models, the mass of such NSs should be around \( 1.25 \, M_{\odot} \). 
Oxygen-neon-magnesium white dwarfs accrete material until they reach the Chandrasekhar mass limit, collapsing into NSs via electron capture \citep{nomoto1984evolution}. 
Accretion-induced collapse may be an important channel for the formation of NSs, especially millisecond pulsars \citep{tauris2013evolution, wang2022formation}. 
Due to accretion, the oxygen-neon-magnesium white dwarf spins up, and accretion-induced collapse may occur only when the mass exceeds the Chandrasekhar limit. 
\citet{yoon2005evolution} found that the Chandrasekhar mass for rapidly rotating white dwarfs can theoretically exceed \( 2.0 \, M_{\odot} \). 
However, more detailed and comprehensive studies are lacking. 

In binary systems, the mass of NSs may increase by accreting matter from their companions. The existence of millisecond pulsars is direct evidence for  matter accretion.The birth spin period of NSs is about \( 50 \, \text{ms} \) \citep{du2402initial}. NSs spin down due to radio radiation, but they can be spun up again to millisecond periods through accretion, a process known as recycling \citep{alpar1982new, radhakrishnan1982origin, bhattacharya1991formation}. 
The discovery of SAX J1808.4–3658 provided direct evidence for the recycling theory \citep{wijnands1998millisecond}. 
To date, the number of discovered accreting millisecond X-ray pulsars (AMXPs) has reached 25 \citep{papitto2022millisecond, sharma2023astrosat}. 
Calculating the recycling process can be used to constrain the birth mass of NSs \citep{you2025determination, tauris2011formation, cognard2017massive}. 
The recycling process is affected by many uncertainties, such as accretion disk instabilities \citep{lasota2001disc} and the propeller effect \citep{romanova2018properties}, making it difficult to accurately compute the mass accreted during this phase.
In some earlier studies of NS accretion, NS mass growth was limited only by the Eddington limit \citep{tauris1999formation}, which may overestimate the accreted mass. 
The mass accreted by a NS is correlated with its spin period; if only the spin-up phase is considered, the spin period and accreted mass are inversely correlated \citep{tauris2012formation}. 
\citet{li2021maximum} calculated the maximum accreted mass by a NS under spin evolution. 
During mass transfer, a large amount of material is ejected due to the propeller effect, which would result in the accreted mass being less than that considering only Eddington accretion. 
Moreover, in  Intermediate/Low-Mass X-ray Binary (I/LMXB) system models, since the mass transfer rate exceeds 100 times the Eddington limit, most of the material is blown away during mass transfer. 

However, the above models did not consider the possibility of super-Eddington accretion. In fact, there are some NSs accreting matter with super-Eddington rates.\citet{bachetti2014ultraluminous} discovered the first pulsating ultraluminous X-ray source. subsequently, several ultraluminous X-ray pulsars have been discovered, including NGC5907 ULX-1 \citep{israel2017accreting}, M51 ULX-7 \citep{castillo2020discovery}, NGC7793 P13 \citep{furst2016discovery, furst2018tale, israel2017discovery}, NGC300 ULX-1 \citep{carpano2018discovery}, SMC X-3 \citep{tsygankov2016propeller, townsend20172016}, NGC2403 ULX \citep{trudolyubov2007chandra}, Swift J0234.6+6124 \citep{doroshenko2018orbit, van2018evolving}, and RX J0209.6-7427 \citep{chandra2020study}. 
The accretors in these systems are believed to be NSs, meaning that NSs can accrete at super-Eddington rates. 
On the spherization of the accretion disk around a neutron star and the interaction between the neutron star’s magnetic field and the accretion disk, \citet{lipunov1982supercritical} conducted a detailed study of super-Eddington accretion onto magnetized NSs and summarized it in \citet{lipunov1987ecology}. 
\citet{lu2012population, zhu2013low} used this theory to simulate the population of symbiotic X-ray binaries (SyXBs). 
Both observations and theory support the possibility of super-Eddington accretion, indicating that pulsars may gain more mass than previously expected.

The present paper focuses on the mass change of NSs with  super-Eddington accretion, and finally uses population synthesis to calculate the possible mass distribution and spin period distribution of NSs.The structure of this paper is as follows: Section 2 describes the model and relevant details. Section 3 presents the evolution of NSs interacting with companions through specific cases and provides results from a large-sample study. Section 4 discusses uncertainties in the model and provides the summary of this study.


\section{Model Inputs and Methods}
In binary evolution models, the common envelope evolution and the effects of supernovae remain highly uncertain \citep{han2020binary,wang2025neutrino,luo2025effect,he2024impact}. Therefore, we employ two codes to simulate binary evolution involving NSs with super-Eddington accretion. First, by utilizing the BSE code developed by \citet{hurley2002evolution}, we can rapidly obtain a set of binary systems consisting of a NS and a main-sequence companion. Then we employ the Modules for Experiments in Stellar Astrophysics (MESA, version r24.08.1;\citet{paxton2013modules,paxton2015modules,paxton2018modules,paxton2019modules}) to perform detailed calculations for the accretion process onto the NS from its main-sequence companion under stable mass-transfer conditions. Finally, we apply bilinear interpolation to map the NS binary distribution generated by BSE onto the MESA grid.
\subsection{ Formation of NS binary }
We use BSE code and population synthesis method to simulate the formation of NS binaries. The input parameters and key physical assumptions are briefly described as follows. We employ a Monte Carlo approach to evolve $10^7$ primordial binary systems. In the binary population synthesis method, for initial primary masses $M \in [5, 100]~M_{\odot}$, we use the initial mass function $\phi(M) \propto M^{-2.3}$ \citep{kroupa2001variation}. The mass ratio ($q$) follows a uniform distribution in the range \([0, 1]\) ; the initial eccentricity distribution follows $\chi(e) \propto e^{-0.5}$ with $e \in [10^{-5}, 0.9]$ \citep{sana2012binary}. The initial orbital period distribution is given by $\psi(\log(P/\text{days})) \!\propto\! \left(\log(P/\text{days})\right)^{-0.55}$ where $\log(P/\text{days})\in [0.15, 4.0]$ \citep{sana2012binary,pauli2022synthetic}. the NS natal kick velocities follow a Maxwellian distribution with a dispersion of $265~\text{km/s}$ \citep{hobbs2005statistical}. The initial metallicity for all stars is $Z = 0.02$ with an initial hydrogen abundance of $X = 0.7$. The criterion for stable mass transfer follows the scheme of \citet{hjellming1987thresholds,hurley2002evolution}. If unstable mass transfer occurs, the common envelope evolution is treated using \citet{tout1997rapid} with a common envelope efficiency $\alpha_{\mathrm{CE}}=1.0$. The orbital periods and initial companion mass distributions of the final binary system sample are constrained to be consistent with the screening range of \citet{li2021maximum}, as shown in the figure below.
\begin{figure}
	\centering
	\includegraphics[width=\linewidth]{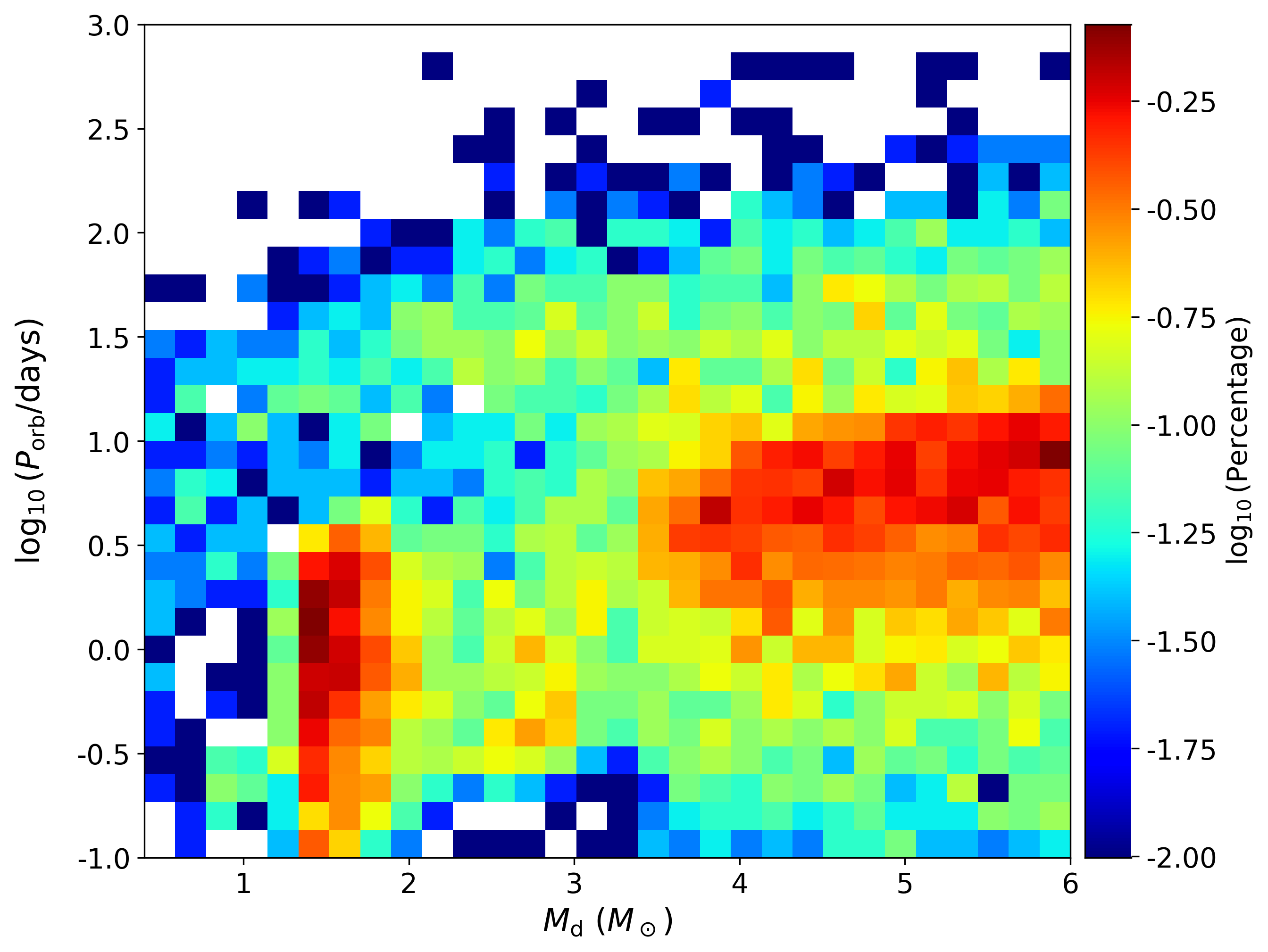}
	\caption{The distribution of orbital periods and companion masses of NS binaries. The color scale represents the logarithm of the percentage of systems in each grid cell.		
		\label{fig1}}
\end{figure}
For systems with an initial companion mass less than $1\,M_{\odot}$, stable mass transfer tends to drive them to evolve into Redback and Black Widow pulsar binaries \citep{chen2013formation, roberts2013neutron}. When the initial companion mass exceeds $4.88\,M_{\odot}$, the mass transfer rate can easily surpass $10^{4}$ times the Eddington limit. Even when super-Eddington accretion is considered, this would result in over 95\% of the material being ejected from the binary system. Therefore, when the mass transfer rate exceeds $10^{4}$ times the Eddington limit, we assume that the mass transfer in such systems is unstable. Similarly, for systems with longer orbital periods, the companion is more likely to overflow its Roche lobe during the giant phase, making the system more prone to entering an unstable state.
 
\subsection{ Evolution of NS binaries }
By the BSE code and the population synthesis method, we obtain 14198 NS binaries from the $10^7$ primordial binary systems we simulated. Figure 1 shows the distribution of orbit periods and the companion masses of 14198 NS binaries. In order to carefully simulate the evolution of NS binaries,  we use  MESA. During evolution, the companion may transfer mass to the NS via Roche lobe overflow. This process can lead to unstable mass transfer, and the binary may undergo common envelope evolution. Since the amount of mass that a NS can accrete during common envelope evolution remains uncertain, we focus primarily on calculating accretion under stable mass transfer via Roche lobe overflow.

In the simulation, the NS is treated as a point mass. The mixing length parameter is set to $\alpha_{\text{MLT}} = 1.9$. The mass transfer scheme follows \citet{ritter1988turning}:
\begin{equation}
\dot{M} \propto \frac{R_{\mathrm{RL,d}}^{3}}{M_{\mathrm{d}}} \exp\left(\frac{R_{\mathrm{d}} - R_{\mathrm{RL,d}}}{H_{\mathrm{p}}}\right),
\label{eq:mass_loss}
\end{equation}
where $R_{\mathrm{d}}$ is the radius of the secondary star, $R_{\mathrm{RL,d}}$ is the Roche lobe radius, $M_{\mathrm{d}}$ is the mass of the secondary star, and $H_{\mathrm{p}}$ is the pressure scale height. In MESA, orbital angular momentum loss is primarily classified into three types.

\textbf{1. Gravitational braking}

The orbital angular momentum carried away by the gravitational-wave radiation can be calculated as 
\begin{equation}
\dot{J}{\mathrm{gr}} = -\frac{32}{5} \frac{\mathrm{G}^{3}}{\mathrm{c}^{5}} \frac{M_{\mathrm{a}} M_{\mathrm{d}} (M_{\mathrm{a}} + M_{\mathrm{d}})}{a^{4}} J,
\end{equation}
where $\mathrm{G}$ is the gravitational constant, $\mathrm{c}$ is the speed of light, $M_{\mathrm{a}}$ is the mass of the accreting star, $J$ is the total angular momentum, and $a$ is the semi-major axis of the binary system.

\textbf{2. Magnetic braking}

Magnetic braking converts the star's rotational angular momentum into outflowing angular momentum carried by the stellar wind through the coupling between the stellar magnetic field and the wind material. In tidally locked binary systems, the loss of rotational angular momentum effectively extracts orbital angular momentum from the system via tidal coupling. This leads to continuous orbital contraction and can trigger or enhance stable mass transfer between the binary components.In MESA, the angular momentum loss due to magnetic braking can be expressed as
\begin{equation}
\dot{J}_{\text{mb}} = -3.8 \times 10^{-30} \cdot M_{\mathrm{d}} \cdot R_{\odot}^4 \cdot \left( \frac{\min(R_{\mathrm{d}}, R_L)}{R_{\odot}} \right)^{\gamma_{\mathrm{mb}}} \cdot \left( \frac{2\pi}{P_{\text{orb}}} \right)^3 \cdot S
\end{equation}
The formula describes the orbital angular momentum loss rate $\dot{J}_{\mathrm{mb}}$ due to magnetic braking in a binary system. Here, $M$ denotes the mass of the donor star; $R_{\odot}$ is the solar radius constant; $\min(R_{\mathrm{d}}, R_{\mathrm{L}})$ represents the smaller value between the star's actual radius $R_{\mathrm{d}}$ and its Roche lobe radius $R_{\mathrm{L}}$; $P_{\text{orb}}$ is the orbital period of the binary; $S$ is a scaling factor that physically adjusts the strength of magnetic braking based on the mass fraction of the star's convective envelope; and $\gamma_{\mathrm{mb}}$ is the magnetic braking index parameter , and $\gamma_{\mathrm{mb}}$ is assigned a value of 4 following the standard magnetic braking prescription \citep{chen2013formation, van2019low, zhou2026evolutioncataclysmicvariablesdifferent}.

\textbf{3. Angular momentum loss due to mass loss}

During the mass transfer process, we assume that all material transferred from the companion star flows toward the NS. We define $f_{\mathrm{mt}}$ as the fraction of matter transferred from the companion that is accreted by the neutron star. The fraction of matter blown away from the binary system due to radiation pressure from the accretion disk near the neutron star is denoted as $\beta_{\mathrm{mt},\beta}$, while the fraction $\beta_{\mathrm{mt},\delta}$ forms a circumbinary disk. Consequently, we have $\beta_{\mathrm{mt},\beta} = 1 - \beta_{\mathrm{mt},\delta} - f_{\mathrm{mt}}$. The accretion efficiency will be discussed in detail in Section 2.2.

The angular momentum carried away by the material lost near the NS is
\begin{equation}
\dot{J}{\mathrm{ml},\mathrm{\beta}} = \beta{\mathrm{mt},\mathrm{\beta}} \dot{M} \left( \frac{M_{\mathrm{d}}}{M_{\mathrm{a}} + M_{\mathrm{d}}} a \right)^{2} \omega \sqrt{1 - e^{2}},
\end{equation}
where $\dot{M}$ denotes the mass transfer rate, $a$ represents the semi-major axis, and $e$ is the eccentricity of the binary system.
Additionally, a portion of the material may form a circumbinary disk during the mass transfer process
\begin{equation}
\dot{J}{\mathrm{ml},\mathrm{\delta}} = \beta{\mathrm{mt},\mathrm{\delta}} \dot{M} \gamma_{\mathrm{disk}} \sqrt{\mathrm{G} (M_{\mathrm{a}} + M_{\mathrm{d}}) a}.
\end{equation}
Here, according to observational data, $\gamma_{\mathrm{disk}}^{2} = 1.7$ \citep{muno2006mid}, and $\beta_{\mathrm{mt},\mathrm{\delta}} = 3 \times 10^{-4}$ \citep{taam2003cataclysmic}. For simplicity, the initial neutron star mass is set to $1.4~M_\odot$. The initial secondary mass is constrained to the range $0.4\,M_{\odot} \leq M_{\mathrm{d}} \leq 6.0\,M_{\odot}$, and we adopt a uniform mass step of $\Delta M_{\mathrm{d}} = 0.28\,M_{\odot}$. We restrict initial orbital periods to the range $0.1\,\text{d} \leq P_{\mathrm{orb}} \leq 316.2\,\text{d}$ ($10^{-1}\,\text{d}$ to $10^{2.5}\,\text{d}$), employing a uniform logarithmic step size of $\Delta \log_{10}(P_{\mathrm{orb}}/\text{days}) = 0.1$. The evolutionary calculations are terminated at the Hubble time. Mass transfer is considered to cease when $\dot{M} \leq 10^{-12}\,M_{\odot}\,\text{yr}^{-1}$ \citep{chen2017formation}.
\subsection{ Evolution of accreting NSs }
The interaction between rotating magnetized NSs and ambient matter has been widely recognized by the academic community as one of the most significant astrophysical problems in NS evolution \citep{pringle1989accretion,illarionov1975number,ghosh1978disk,lovelace1995spin,lovelace1999magnetic}. Generally, the accretion onto NSs is constrained by the Eddington limit, that is
\begin{equation}
\dot{M}_{\text{Edd}} =3.6 \times 10^{-8} \left( \frac{M_{\mathrm{NS}}}{1.4 M_{\odot}} \right) \left( \frac{0.1}{\mathrm{G} M_{\mathrm{NS}} / R_{\mathrm{NS}} \mathrm{c}^{2}} \right) \left( \frac{1.7}{1 + X} \right) M_{\odot} \mathrm{yr}^{-1}.
\end{equation}
The interaction between a rotating magnetized NS and its environment can be characterized through the relationships among three characteristic radii, where $M_{\mathrm{NS}}$ represents the NS mass, $\mathrm{c}$ denotes the speed of light, and $R_{\mathrm{NS}}$ indicates the NS radius, $R_{\mathrm{NS}} = 15 (M_{\mathrm{NS}}/M_{\odot})^{-1/3} \, \mathrm{km}$ \citep{tauris2012formation}.

The light cylinder radius $R_{\mathrm{l}}=\mathrm{c}/\omega_{\mathrm{spin}}$, where $\omega_{\mathrm{spin}}$ is the spin angular velocity of the NS. At the corotation radius $R_{\mathrm{c}}$, the rotational velocity of the NS magnetosphere matches the orbital velocity of accreting matter in the Keplerian disk. It is defined as: $ R_{\mathrm{c}} = \left( \frac{\mathrm{G} M}{\omega_{\mathrm{spin}}^{2}} \right)^{1/3} $. The magnetospheric radius $R_{\mathrm{mag}}$ is where the pressure of accreting matter equals the magnetic pressure of the NS's magnetic field. The NS's magnetic field disrupts the accretion disk structure at this radius, preventing accreting matter from following Keplerian orbital motion. Due to magnetic flux freezing, plasma at this radius is forced to co-rotate with the NS.    
We assume that the magnetospheric radius equals the Alfv\'{e}n radius, i.e., $R_{\mathrm{mag}} = R_{\mathrm{a}}$. The Alfv\'{e}n radius is expressed by the following formula
\begin{equation}
R_{\mathrm{a}} = \left( \frac{\mu^{2}}{\dot{M}_{\mathrm{NS}} \sqrt{2 \mathrm{G} M_{\mathrm{NS}}}} \right)^{2/7},
\end{equation}
where $\dot{M}_{\mathrm{NS}}$ is the accretion rate of the NS, and $\mu = B_{\mathrm{NS}} R_{\mathrm{NS}}^3 / 2$ is the magnetic dipole moment, with $B_{\mathrm{NS}}$ being the magnetic field of the NS. After the NS accretes matter, the magnetic field decays, and the decay formula is \citep{oslowski2011population}
\begin{equation}
B_{\mathrm{NS}} = \left( B_{\mathrm{NS}}^{\mathrm{i}} - B_{\mathrm{min}} \right) \exp \left( -\frac{\Delta M}{M_{\mathrm{B}}} \right) + B_{\mathrm{min}},
\end{equation}
where $B_{\mathrm{NS}}^{\mathrm{i}} = 5 \times 10^{12} \, \mathrm{G}$ is the initial magnetic field strength,we assume the minimum magnetic field is $5 \times 10^{8}$ G.$\Delta M$ is the mass accreted by the NS, $M_{\mathrm{B}}$ is set to $0.025 \, M_{\odot}$. The uncertainties associated with the magnetic field will be addressed in Section 4. Disk evolution under super-Eddington mass transfer rate, \citet{shakura1973black} suggested that the accretion rate in the disk decreases monotonically from a certain radius $R = R_{\mathrm{s}}$ (the spherization radius). Regarding the super-Eddington accretion process on magnetized NSs, it is summarized in \citet{lipunov1987ecology}. The sphericalization radius \( R_{\mathrm{s}} \) can be approximately expressed as
\begin{equation}
R_{\mathrm{s}} = \frac{\kappa}{4\pi \mathrm{c}} \dot{M},
\end{equation}
where $\kappa$ represents the opacity of the accreted matter. When $R_{\mathrm{a}} > R_{\mathrm{s}}$, the NS accretion rate $\dot{M}_{\mathrm{NS}} = \dot{M}$. When $R_{\mathrm{a}} < R_{\mathrm{s}}$, the accretion rate becomes $\dot{M}_{\mathrm{NS}} = \frac{R_{\mathrm{a}}}{R_{\mathrm{s}}} \dot{M}$. By setting $R_{\mathrm{s}} = R_{\mathrm{a}}$, we obtain a critical condition for $\dot{M}_{\mathrm{NS}}$, denoted as $\dot{M}_{\mathrm{crit}}$ and expressed as
\begin{equation}
\dot{M}_{\mathrm{crit}} \approx \left( \frac{\mu^{2}}{\sqrt{2\mathrm{G}M}} \right)^{\frac{2}{9}} \left( \frac{4\pi \mathrm{c}}{\kappa} \right)^{\frac{7}{9}} \propto \mu^{4/9}.
\end{equation}  
The excess material is blown out of the binary system by radiation pressure, and we assume that the loss occurs in the vicinity of the neutron star, the Alfv\'en radius can be expressed as
\begin{equation}
R_{\mathrm{Sa}}=\left(\frac{\mu^{2}\kappa}{4\pi \mathrm{c}\sqrt{2 \mathrm{G} M}}\right)^{2/9}.
\end{equation}

Based on the inter-relations among these three radii, NSs may exist in several basic evolutionary states \citep{lipunov1992astrophysics}.

\textbf{Radio phase} ($R_{\mathrm{a}} > R_{\mathrm{l}}$): The NS acts as a radio pulsar, spinning down due to magnetic dipole radiation. The spin-down torque can be expressed as
\begin{equation}
K_{\mathrm{sd}} = \frac{2\mu^{2}}{3R_{\mathrm{l}}^{3}}.
\end{equation}
This formula should include an angle $\xi$ between the rotation axis and the magnetic axis. The long-term variation of $\xi$ is not considered in our study. During this process, the NS does not accrete matter, so we set $f_{\mathrm{mt}} = 0$ at this stage.

\textbf{Propeller phase} ($R_{\mathrm{a}} < R_{\mathrm{l}}$, $R_{\mathrm{a}} > R_{\mathrm{c}}$): Since the disk is truncated beyond the corotation radius, the material velocity at the magnetically frozen Keplerian disk is less than the rotational velocity of the NS's magnetic field. The material flow exerts a spin-down effect on the NS's rotation. The torque can be expressed as
\begin{equation}
K_{\mathrm{sd}} = \frac{k_{\mathrm{t}} \mu^{2}}{R_{\mathrm{a}}^{3}}.
\end{equation}
The numerical factor $k_{\mathrm{t}}$ depends on the specific model, but its value is approximately 1 \citep{shakura2012theory}. Due to the acceleration of accreted matter by the magnetic field, the gravitational force acting on the matter becomes less than the centrifugal force, which prevents the matter from being accreted by the NS. Therefore, $f_{\mathrm{mt}} = 0$.

\textbf{Accretion phase} ($R_{\mathrm{a}} < R_{\mathrm{c}}$): When the magnetospheric radius is smaller than the corotation radius, the angular momentum carried by the matter will be transferred to the NS, and the material will then follow the magnetic field lines to accrete onto the NS's magnetic poles. On the other hand, since the accretion disk interacts with the NS's magnetic field, the interaction between the NS's magnetic field and the accretion disk needs to be divided into two components. The net torque can be expressed as
\begin{equation}
K = \dot{M}_{\mathrm{NS}} \sqrt{\mathrm{G} M_{\mathrm{NS}} R_{\mathrm{a}}} - \frac{1}{3} \mu^{2} R_{\mathrm{c}}^{-3}.
\end{equation}
Based on the discussion of Equation (10) and considering the uncertainties in the NS accretion process, we assume an accretion efficiency of 0.5 \citep{he2024constraining}, assuming that the mass loss occurs in the vicinity of the neutron star and carries away its specific orbital angular momentum. For convenience, we denote the accretion efficiency as $f_{\mathrm{mt},\mathrm{a}}$, which is expressed as
\begin{equation}
f_{\mathrm{mt},\mathrm{a}} = 
\begin{cases}
0.5 (1 - \beta_{\mathrm{mt},\mathrm{\delta}}), &\quad \dot{M} < \dot{M}_{\mathrm{crit}}, \\
0.5 (1 - \beta_{\mathrm{mt},\mathrm{\delta}}) \cdot \dfrac{R_{\mathrm{a}}}{R_{\mathrm{s}}}, &\quad \dot{M} > \dot{M}_{\mathrm{crit}}.
\end{cases}
\end{equation}
When the NS is in the subcritical accretion phase, its spin equilibrium period is determined by the following formula \citep{lipunov1992astrophysics}
\begin{equation}
P^{\mathrm{eq}} = 5.72 M_{\mathrm{NS}}^{-5/7} \dot{M}_{16}^{-3/7} \mu_{30}^{6/7} \, \mathrm{s}.
\end{equation}
When the NS is in the supercritical accretion phase, its spin equilibrium period is given by
\begin{equation}
P_{\mathrm{s}}^{\mathrm{eq}} = 1.76 \times 10^{-1} \mu_{30}^{2/3} M_{\mathrm{NS}}^{-2/3} \, \mathrm{s}.
\end{equation}

When $R_{\mathrm{c}} \sim R_{\mathrm{a}}$, we also consider that the NS enters an equilibrium state. At this point, the accretion disk exerts no torque on the NS, and we assume the accretion efficiency remains consistent with that during accretion.

Finally, the value of $f_{\mathrm{mt}}$ under various states is briefly summarized and expressed as
\begin{equation}
f_{\mathrm{mt}} = 
\begin{cases}
0, & \text{propeller and radio phases}, \\
f_{\mathrm{mt},\mathrm{a}}, & \text{equilibrium and accretion states}.
\end{cases}
\end{equation}

Since we have not considered the influence of the pulsar's equation of state on accretion, we simply set the moment of inertia of the pulsar to $I = 10^{45} \ \mathrm{g\cdot cm^{2}}$.

\section{Result}
\label{sect:data}
Regarding NS accretion, numerous studies have been conducted by previous researchers. Among them, \citet{li2021maximum} did not consider super-Eddington accretion in NSs, while our model incorporates super-Eddington accretion. To facilitate the research and identify the differences between the two models, we have reproduced the study by \citet{li2021maximum} and compared it with our own model.
\subsection{ Evolutionary Examples }
Figure 2 presents the evolution of mass-transfer rate,
NS mass, Alfven radius and corotation radius in a 
binary system calculated by the model without the super
-Eddington accretion from Li et al. (2021), which 
closely resembles the right panel of Figure 2 in their 
paper. The companion star fills its Roche lobe within the Hertzsprung gap. During the initial phase of mass transfer, which occurs on a thermal timescale, the mass transfer rate increases rapidly. In the model of \citet{li2021maximum}, the primary growth in the NS's mass originates from this stage. The upper limit of accretion is constrained by the Eddington limit. After the NS accretes a small amount of material , the mass transfer rate \(\dot{M}\) reaches its peak, it gradually declines as the source parameters evolve, the propeller effect begins to operate, where the magnetosphere accelerates the accreting matter, ejecting it from the binary system, as the NS's rotation slows down, the corotation radius gradually expands. However, as it approaches the magnetospheric radius, the spin-down torque will decrease. This will cause the system to remain in a long-term propeller regime when the mass transfer rate drops. Therefore, in this model, the system is constrained by the Eddington limit and resides in a prolonged propeller phase, resulting in relatively little accretion onto the NS.
\begin{figure*}
	\centering
	\includegraphics[width=\linewidth]{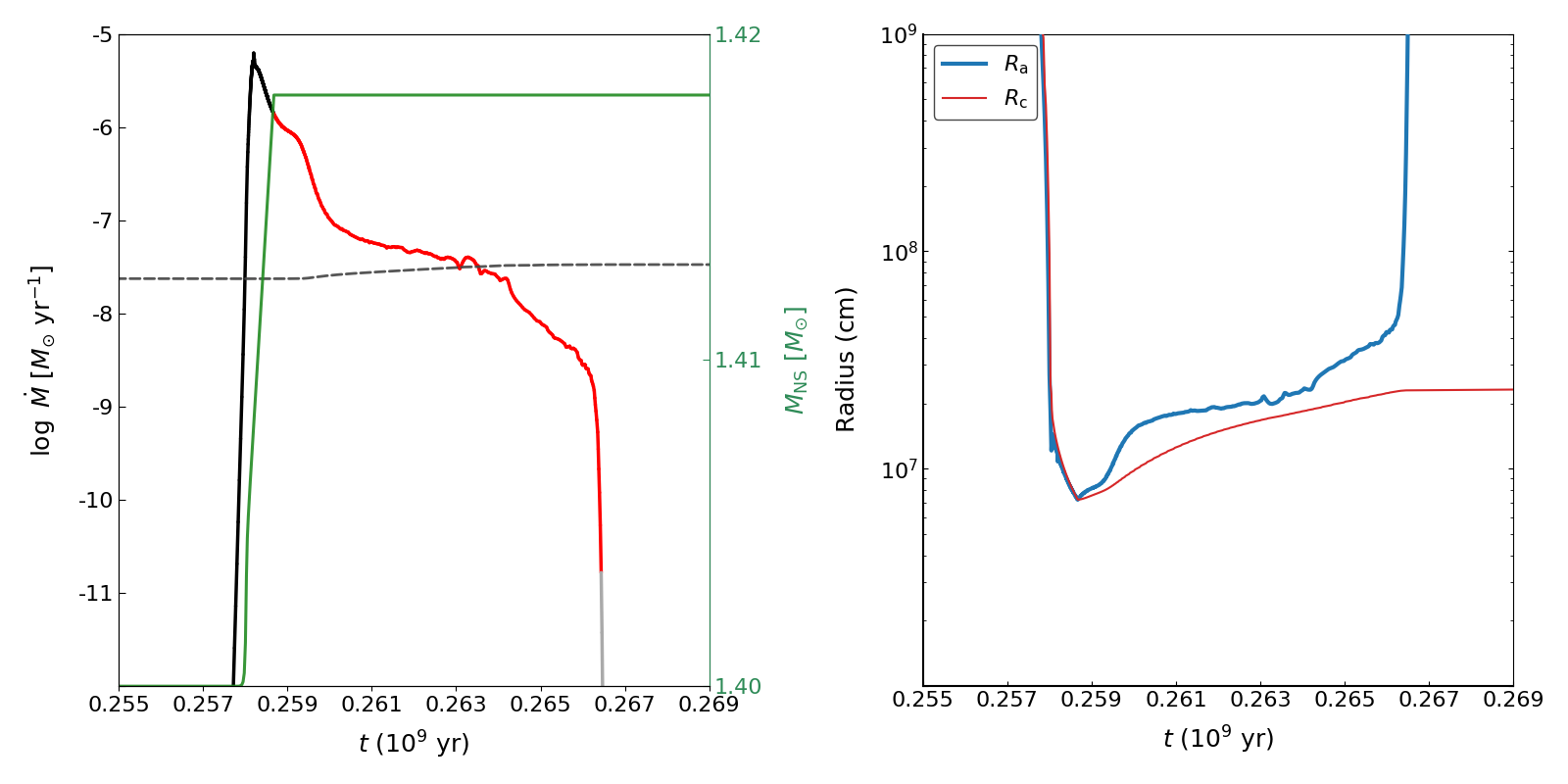}
	\caption{
		The left panel shows the mass transfer rate \(\dot{M}\) (black line) and the NS mass \(M_{\mathrm{NS}}\) (green line) as functions of stellar age for an initial NS mass \(M_{\mathrm{NS,i}} = 1.4\,M_{\odot}\), an initial donor mass \(M_{\mathrm{d,i}} = 3.2\,M_{\odot}\), and an initial orbital period \(P_{\mathrm{orb,i}} = 1.6\,\mathrm{d}\). The region where the propeller (ejection) effect occurs is highlighted in red, and the Eddington accretion rate \(\dot{M}_{\mathrm{Edd}}\) is indicated by the grey dashed line. The right panel shows the evolution of the magnetospheric radius \(R_{\mathrm{m}}\) (blue line) and the corotation radius \(R_{\mathrm{co}}\) (red line).
	}
	\label{fig2.png}  
\end{figure*}
\begin{figure*}
	\centering
	\includegraphics[width=\linewidth]{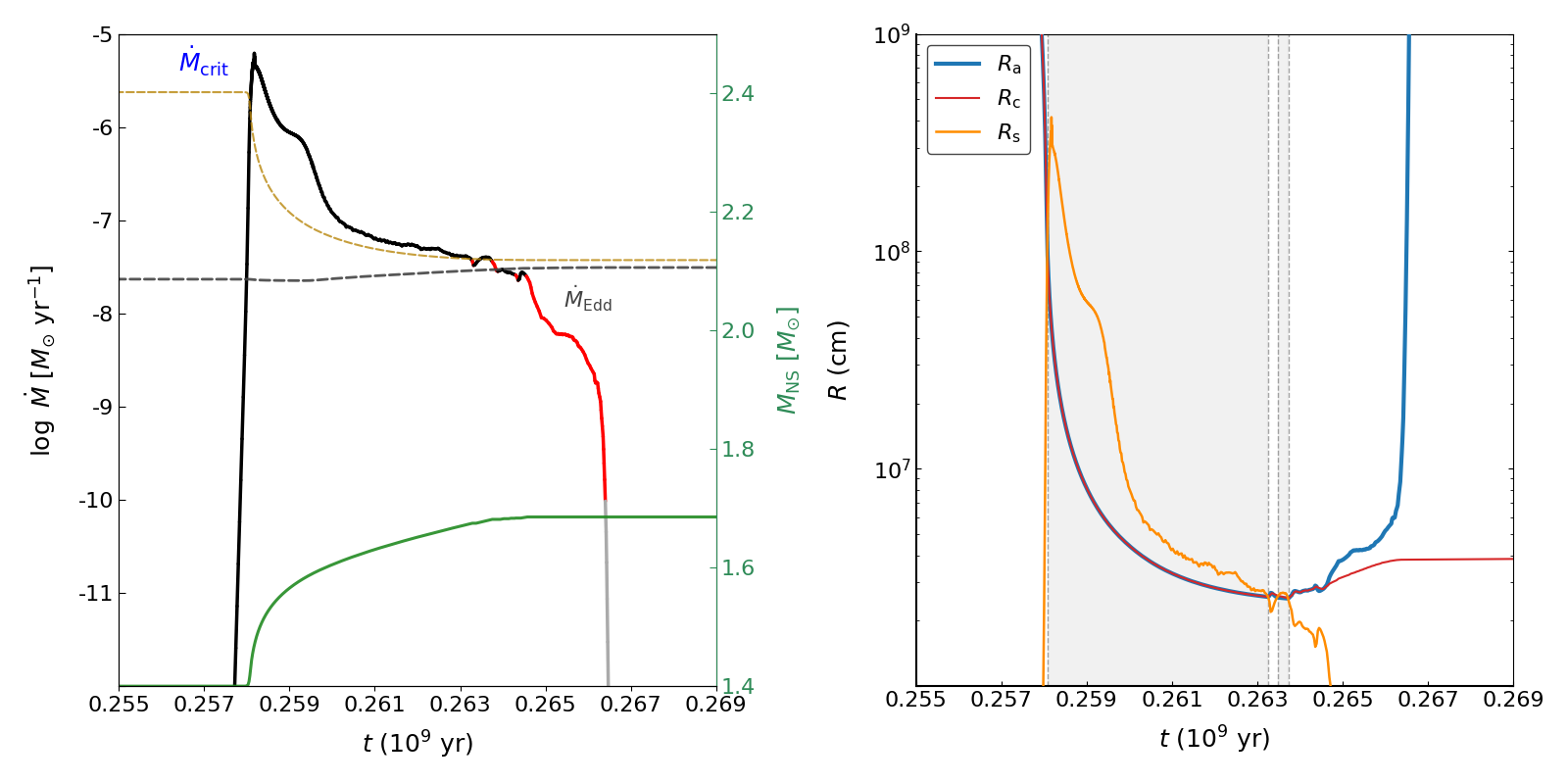}
	\caption{
		Similar to Figure 2, but for the our model with super-Eddington accretion. In the left panel, the orange dashed line indicates the critical mass accretion rate \(\dot{M}_{\mathrm{crit}}\) for supercritical accretion. The gray solid line denotes when the system enters the radio pulsar phase. In the right panel, the orange line represents the spherization radius for NS accretion, and the gray shaded region indicates the occurrence of supercritical accretion. Further details can be found in the main text.
	}
	\label{fig3.png}  
\end{figure*}

Figure 3 illustrates the results of our model considering super-Eddington accretion. In this model, when the mass transfer rate exceeds the critical value $\dot{M}_{\mathrm{crit}}$, the system enters a supercritical accretion regime. The accretion efficiency can be described by Equation~(14), with the excess material being lost from the vicinity of the neutron star, carrying away angular momentum. The pressure resisting the accretion flow transitions from being dominated by magnetic pressure to radiation pressure. The scale of the magnetospheric radius is no longer related to the mass transfer rate, but instead depends solely on the magnetic field strength, i.e., its description transitions from Equation (7) and (11). The magnetic field strength is described by Equation (8), after accreting a certain amount of material, the magnetic field strength approaches its minimum, at which point the magnetospheric radius also gradually declines to its minimum. Meanwhile, the torque acting on the NS is sufficient to maintain \(R_{\mathrm{c}} \sim R_{\mathrm{m}}\), allowing the system to remain in an equilibrium state and sustain accretion over an extended period. This results in a final accreted mass in our model that is significantly greater than that in the model of \citet{li2021maximum}. As the system evolves, when the mass transfer rate decreases and falls below \(\dot{M}_{\mathrm{crit}}\), the dominant pressure resisting the accretion flow shifts back from radiation pressure to magnetic pressure. Subsequently, as the mass transfer rate continues to decrease, the magnetospheric radius gradually expands, while the spin-down torque becomes insufficient to adequately slow the NS's rotation. Then, these effects prevent the corotation radius from catching up with the magnetospheric radius. Ultimately, the system exits the equilibrium state and enters the long-term propeller regime.The amount of matter ultimately accreted by a NS strongly depends on the mass transfer process. In Figures~2 and~3, the mass transfer rate $\dot{M}$ exceeds the Eddington accretion rate $\dot{M}_{\mathrm{Edd}}$ for most of the time. However, due to magnetic field decay, even when super-Eddington accretion is considered, approximately $2.5\,M_\odot$ of material is still ejected from the binary system. 

For lower-mass donors, although the mass-transfer rate $\dot{M}$ may be lower, the NS can undergo more efficient accretion. Figure 4 shows the evolution of the models with the lower-mass donors. Obviously, in the two models of Figure 4, the accreting NS masses can exceed $2.0 M_\odot$. There is an abrupt drop in $\dot{M}$, which is caused by the discontinuity in the chemical composition gradient during the first dredge-up phase \citep{tauris1999formation, istrate2016models}. After the stripping process, a sufficient envelope remains to sustain nuclear burning, causing the donor star to expand again and undergo subsequent mass transfer \citep{jia2014formation, li2019formation,lu2025impact}.

The increase in mass of an accreting NS is closely related to its spin period. Figure 5 presents the evolution of the spin period of an accreting NS in the model of \citet{li2021maximum} (the upper panels in Figure 5) and in the super-Eddington accretion model of this paper (the lower panels in Figure 5).Obviously, the most significant difference between the two models is that, with the decrease in the spin period, in the former model, the NS primarily remains in the propeller stage, whereas in our model, it is mainly in the accretion phase. In their model, a larger mass increment of the NS, $\Delta M_{\mathrm{NS}}$, generally corresponds to a higher spin velocity. The reason may lie in the fact that in the \citet{li2021maximum} model, the magnitude of the spin-up torque is strongly dependent only on the accretion rate of the NS. This implies that at a given accretion rate, the spin-up torque generated by the accretion process does not decay. Consequently, in the \citet{li2021maximum} model, accretion causes its spin period to rapidly shorten. The decrease in spin period leads to its co-rotation radius becoming smaller than the magnetospheric radius, at which point the NS enters a prolonged propeller regime. The decrease in spin period leads to its co-rotation radius becoming smaller than the magnetospheric radius, at which point the NS enters a prolonged propeller regime.
\begin{figure*}
	\centering
	\includegraphics[width=\linewidth]{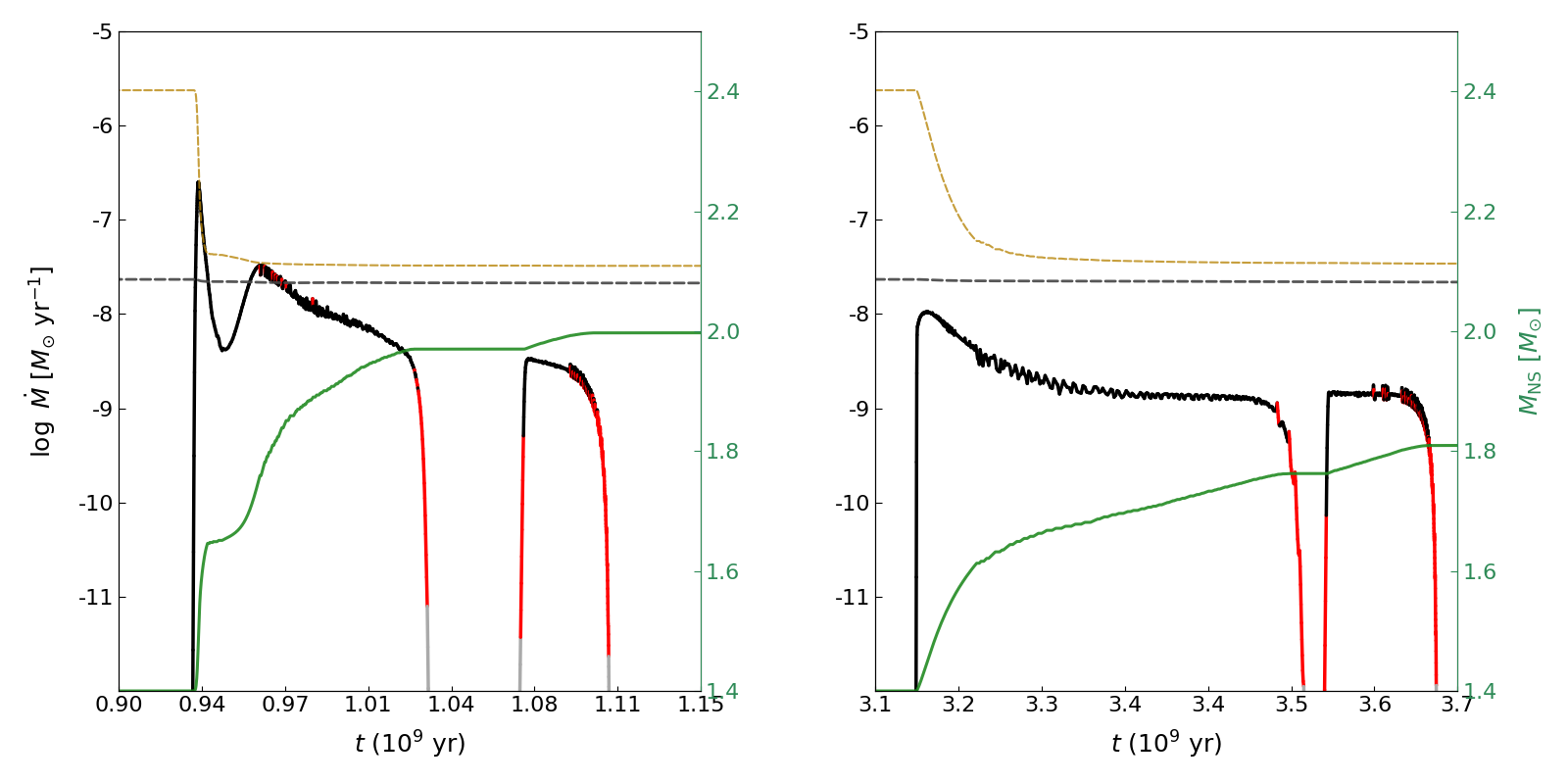}
	\caption{
		Similar to the left panel of Figure~3, but with modified initial parameters for the model: the initial companion star mass and orbital period are set to \(M_{\mathrm{d,i}} = 2.0\,M_{\odot}\), \(P_{\mathrm{orb,i}} = 1.3\,\mathrm{d}\) for the left panel, and to \(M_{\mathrm{d,i}} = 1.4\,M_{\odot}\), \(P_{\mathrm{orb,i}} = 1.6\,\mathrm{d}\) for the right panel. The initial NS mass remains unchanged at \(M_{\mathrm{NS,i}} = 1.4\,M_{\odot}\).
		\label{fig4.png}}
\end{figure*}
\begin{figure*}
	\centering
	\includegraphics[width=\linewidth]{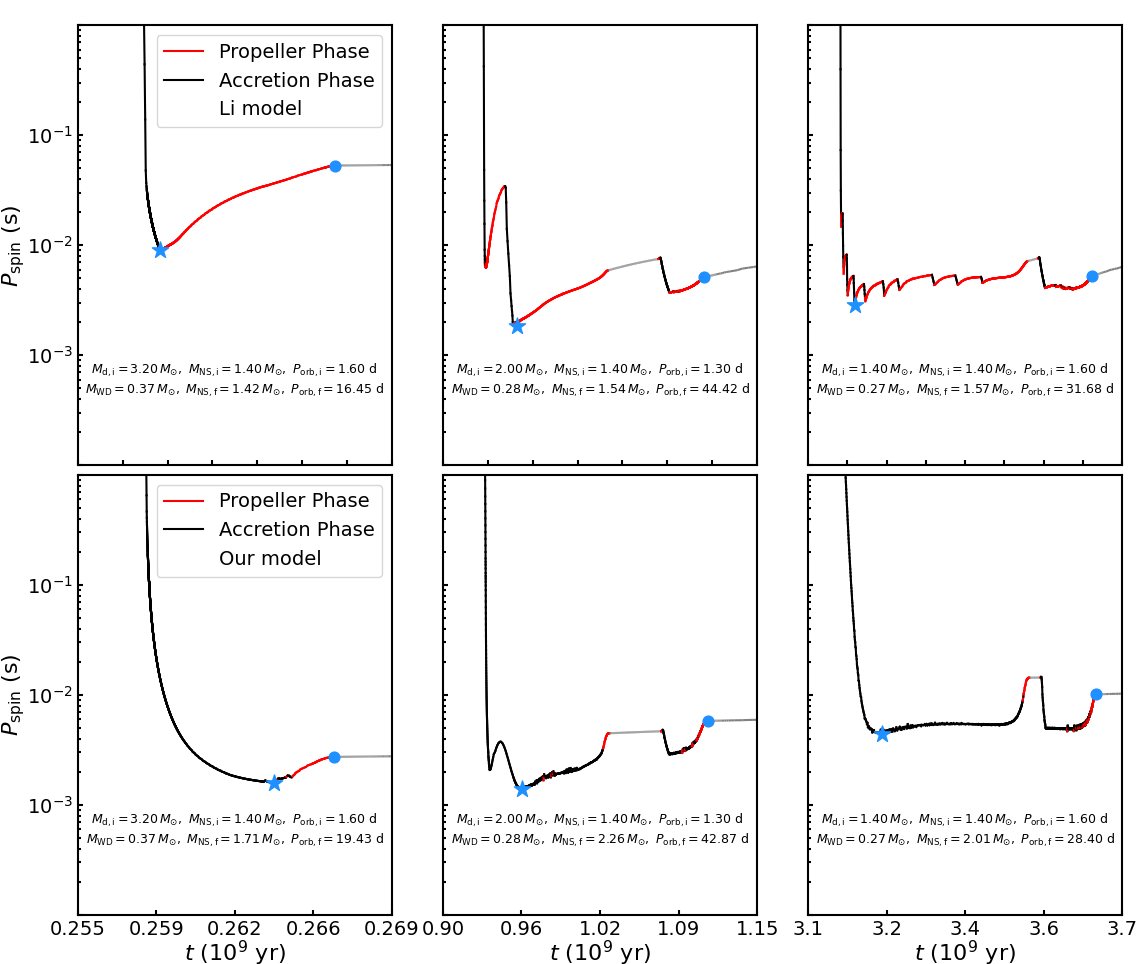}
	\caption{ Spin evolution of NSs with different initial parameters under the model of \citet{li2021maximum} (top row) and our model (bottom row). The initial parameters and the parameters at the end of mass transfer are indicated in the figure. The black segments of the curves indicate periods when the NS is in the accretion process, while the red segments represent phases dominated by the propeller regime. The blue pentagram marks the moment when the spin period reaches the minimum value, and the blue solid circle indicates the time when mass transfer ends.		
		\label{fig5.png}}
\end{figure*}
In our model, as the magnetic field decays or the mass transfer rate changes, the magnetospheric radius will vary. During this process, the torque exerted by the accretion disk on the NS is sufficient to allow the NS's spin rate to keep pace with the evolution of the accretion disk itself, i.e., $R_{\mathrm{c}} \sim R_{\mathrm{a}}$. Since $R_{\mathrm{c}}$ is positively correlated with the NS's spin period, the trend of the spin period evolution in our model will be similar to that of the magnetospheric radius. When the magnetospheric radius evolves to its minimum value, the NS's spin period also reaches its minimum. Therefore, in our model, there is no relationship between the accreted mass and the maximum spin velocity that the NS can achieve.
\subsection{Mass Distribution of NSs in Binary System}
Similar to the work of \citet{gao2022formation}, we first utilized MESA to determine the regions within the parameter space of the initial companion mass $M_{\mathrm{d,i}}$ and the initial orbital period $\log_{10}(P_{\mathrm{orb,i}}/\mathrm{days})$ where stable mass transfer can occur (as illustrated in the left panel of Figure 6). However, unlike the study by \citet{gao2022formation}, our model cannot produce mass-gap black holes. The reason for this is that \citet{gao2022formation} assume a non-decaying magnetic field in their model. Consequently, the strong magnetic field pushes the inner radius of the accretion disk outward, keeping it far away from the spherization radius, which prevents radiation pressure from blowing away the transferred material. As a result, the total accreted mass in their model is significantly larger. When the companion mass is small, mass transfer may not initiate within 14 billion years, and we have excluded systems that experience Roche lobe overflow right at the onset of the evolution. Figure 6 presents the distribution of the final neutron star mass after the cessation of accretion within the parameter space of the initial companion mass $M_{\mathrm{d,i}}$ and the initial orbital period $\log_{10}(P_{\mathrm{orb,i}}/\mathrm{days})$. The missing region in the right panel, compared to the left panel, is due to the fact that during the evolutionary process, the companion mass of the binary systems in this subset decreases to less than $0.05\,M_\odot$, and their orbital periods drop below 1 hour. We assume that these binary systems evolve into black widow binaries, where the companion star is evaporated. At this stage, the neutron star becomes a single star system; therefore, we excluded this subset of binaries when calculating the NS-WD binary systems. Prior to this, however, the binary systems in this region contribute to the neutron star mass distribution observed in X-ray binaries. As shown in Figures 3 and 4, the mass-transfer process significantly affects the amount of mass a neutron star can accrete during the recycling phase. For relatively massive donor stars (e.g., the case with an initial companion mass of $3.2\,M_\odot$ shown in the left panel of Figure 3, roughly corresponding to the range $3.0 \leq M_{\mathrm{d,i}} \leq 5.0\,M_\odot$), the mass-transfer rate exceeds the critical mass accretion rate ($\dot{M}_{\mathrm{crit}}$) for the vast majority of the mass-transfer duration. Because most of the material is blown away by radiation pressure, this results in a low mass-transfer efficiency $f_{\mathrm{mt}}$. The final post-accretion mass typically falls between $1.43\,M_\odot$ and $1.9\,M_\odot$.For intermediate-mass donor stars ($1.6 \leq M_{\mathrm{d,i}} \leq 3.0\,M_\odot$, e.g., the case in the left panel of Figure 4), the system remains in a sub-Eddington accretion state for most of the time. Compared to systems with companion masses exceeding $3.0\,M_\odot$, the duration of mass transfer in such systems is significantly longer. Consequently, the neutron star can still accrete enough material to reach a final mass exceeding $2.1\,M_\odot$. For lower-mass companion stars in our model ($M_{\mathrm{d,i}} \leq 1.6\,M_\odot$, as shown in the right panel of Figure 4), due to the limited total mass available for transfer, the final mass of the neutron star is distributed between $1.8\,M_\odot$ and $2.1\,M_\odot$.

NSs with known masses are primarily found in X-ray 
binaries and pulsar binaries. With the advancement of radio telescopes and high-precision pulsar timing techniques \citep{antoniadis2013massive,demorest2010two}, as well as developments in neutron star mass measurement through X-ray bursts \citet{kusmierek2011mass,lo2013determining}, an increasing number of neutron star masses have been determined. In several studies (see, e.g., \citealt{ozel2012mass, pejcha2012observed, rocha2023mass}), the mass distributions of neutron stars in binary systems were statistically analyzed, and the data were fitted to obtain a bimodal distribution. Their data primarily include three types of binary systems: double neutron star (DNS) systems, neutron star–main sequence star (NS–MS) systems, and neutron star–white dwarf (NS–WD) systems. The observed masses of neutron stars in DNS systems are generally less than \(1.4\,M_{\odot}\). Typically, the formation of double neutron star systems involves a common envelope phase. In the work of \citet{nie2025modeling}, the mass growth of neutron stars prior to the formation of double neutron stars was simulated, and it was found that the maximum accreted mass of a neutron star is approximately 0.04\,\(M_{\odot}\). Stable mass transfer primarily affects the mass distribution of neutron stars in X-ray binaries and NS–WD binary systems. We extracted the masses of neutron stars in X-ray binaries and NS–WD binaries from the statistical data provided in Table A1 of \citet{rocha2023mass}, plotted their mean values as red histograms in the figure, and subsequently performed Bayesian inference on the dataset to obtain the red curves shown in the Figure 7. We then calculated the mass distributions of neutron stars in X-ray binaries and NS–WD binaries under both the \citet{li2021maximum} model and our own model. 

Using the parameter space for stable mass transfer delineated by our MESA grid, we selected 4983 primordial binaries generated by BSE and applied linear interpolation to this sample , obtaining the mass distribution of neutron stars in binary systems.The left two panels of Figure 7 display the mass distributions of neutron stars in X-ray binaries under two different models. The distribution of X-ray binaries is primarily shaped by systems with lower-mass companions (e.g., the region in Figure~6 where the initial companion mass \(M_{\mathrm{d,i}} \lesssim 1.5\,M_{\odot}\)). In such systems, the mass-transfer rate is relatively low but the duration is significantly longer. Although a combined analysis of Figure~1 and Figure~6 indicates a slightly higher number of binary systems in the region where \(1.5\,M_{\odot} \leq M_{\mathrm{d,i}} \) , the left and right panels of Figure~4 reveal that the total mass-transfer duration differs by approximately a factor of four between these cases. For systems with even lower companion masses, this duration becomes even longer. Consequently, the X-ray binary population we observe is likely dominated by systems where the initial companion mass \(M_{\mathrm{d,i}} < 1.5\,M_{\odot}\).

By statistically analyzing all NS--WD systems originating from NS--main sequence star progenitors, we obtained the two right-hand panels of Figure~7. For systems with an orbital period \(P_{\mathrm{orb,i}} > 20\,\mathrm{days}\) and a companion mass \(M_{\mathrm{d,i}} > 5\,M_{\odot}\), we found that their mass-transfer rates mostly exceeded \(10^{4}\) times the Eddington limit. Therefore, we assumed that the neutron star did not accrete matter within this initial parameter range. Using the \textsc{bse} code, we approximately excluded from this parameter space binaries that would evolve into double NSs, those that would merge to form Thorne–Zytkow objects (TZOs) due to insufficient orbital energy to eject the envelope during a common-envelope episode, and the small fraction that would produce a black hole--NS binary. Ultimately, we retained only those binary systems that underwent an NS--MS phase and eventually evolved into NS--WD systems, and normalized them together with the NS--WD systems produced within the initial parameter range.
\begin{figure*}
	\centering
	\includegraphics[width=\linewidth]{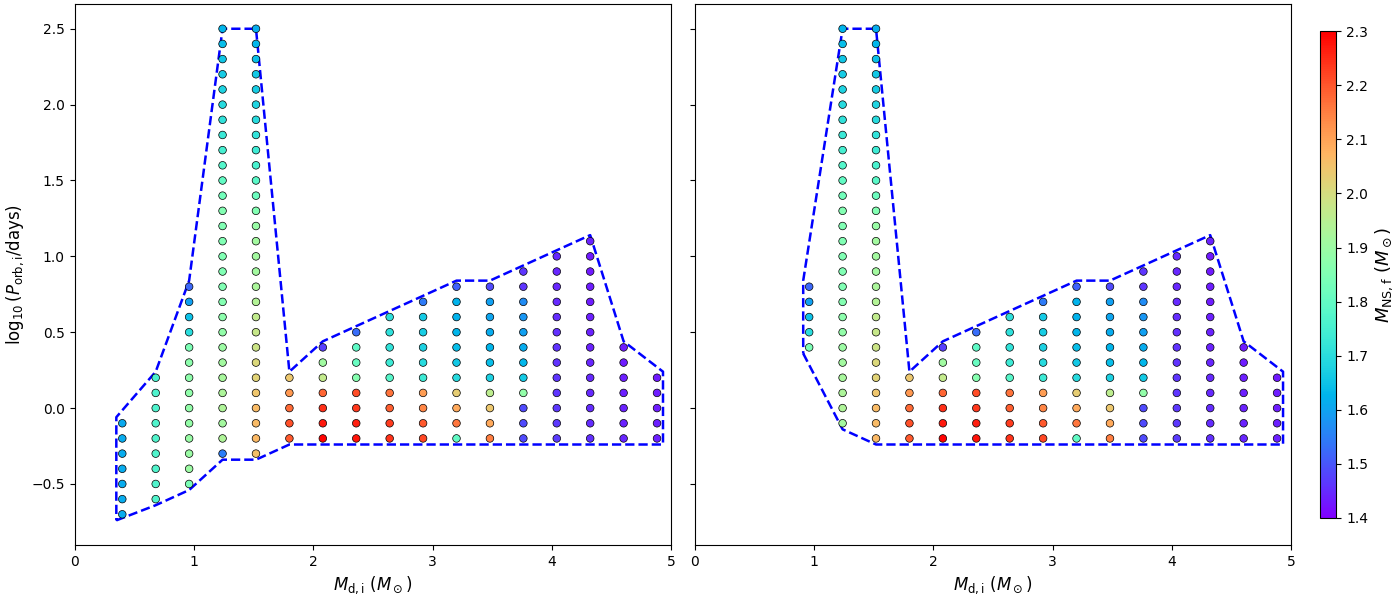}
	\caption{The distribution of the final neutron star mass ($M_{\mathrm{ns,f}}$) within the parameter space of the initial companion mass ($M_{\mathrm{d,i}}$) and the initial orbital period $\log_{10}(P_{\mathrm{orb,i}}/\mathrm{days})$. The left panel displays the entire population of binary systems capable of undergoing stable mass transfer. The right panel displays a subset of these systems, excluding those that evolve into black widow binaries (where the companion mass drops below $0.05\,M_\odot$ and the orbital period is less than 1 hour). The color scale indicates the final mass of the neutron star after accretion ceases. 
		\label{fig6.png}}
\end{figure*}
\begin{figure*}
	\centering
	\includegraphics[width=\linewidth]{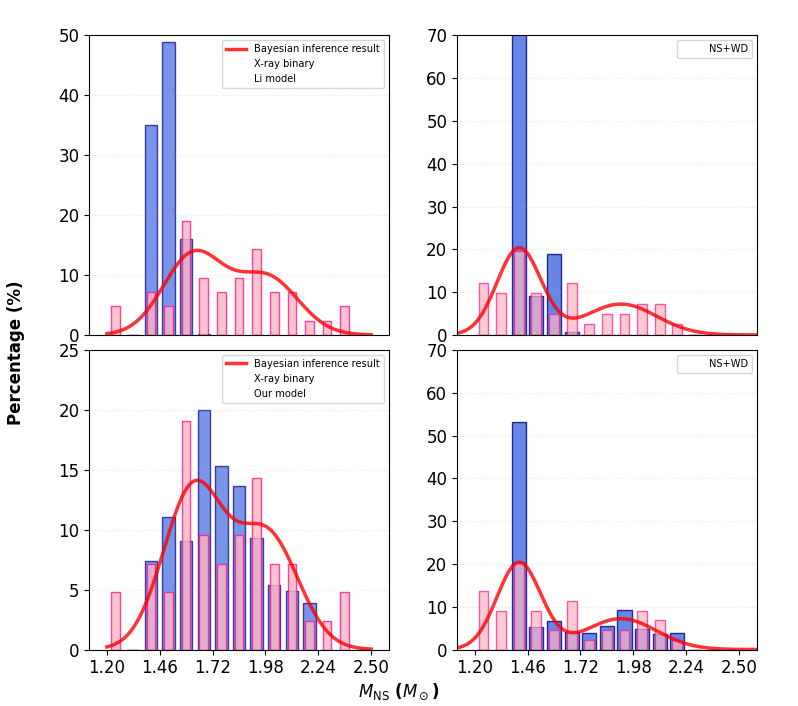}
	\caption{The NS mass distribution in X-ray binaries and NS+WD binaries. In the two panels of the top row, the blue histograms show the NS mass distributions for X-ray binaries and for NS–WD binaries, respectively, under the \citet{li2021maximum} model. In the two panels of the bottom row, the blue histograms show the corresponding distributions for X-ray binaries and for NS–WD binaries in our model. The observational data for the red histogram are sourced from Table A1 of \citet{rocha2023mass} and represent the statistical observational mean values. The red curve displays the result obtained from Bayesian inference.	
		\label{fig7.png}}
\end{figure*}
\subsection{Orbital period and NS spin distribution}
As shown by Figure 5, NS spin is very important for the enhance of accreting NS. \citet{fortin2024catalogue} counted and analyzed the spin peiods and orbital periods for the known LMXB. Figure 8 shows the distribution of NS spin periods vs. orbital periods for the low-mass X-ray binaries. Our result cannot cover the low-mass X-ray bianries with wider orbital periods than $\sim$50 days. They mainly are symbiotic X-ray binaries, which has been investigated by \citet{lu2012population}. It can be observed that the orbital periods are primarily distributed between 0.4 days and 25 days, while the NS spin periods mainly range from 1 ms to 10 ms. During the early phase of thermal-timescale mass transfer, NSs undergo rapid spin-up through accretion, followed by accretion-induced magnetic field decay. Since our magnetic field decay mass scale $M_{\mathrm{B}}$ is set to $0.025 M_\odot$, the magnetic field decays very rapidly. On the other hand, compared to the early thermal-timescale mass transfer phase, the subsequent mass transfer rate is relatively stable. As a result, the Alfvén radius $R_a$ lies between $10^{6}$ cm and $10^{7}$ cm for most of the mass transfer duration. In Section 3.1, we discussed the relationship between NS spin period and $R_a$. Ultimately, the NS spin periods we obtained are roughly concentrated between 1 ms and 15 ms, slightly larger than the observational data. The reason may be that in our model, we set the minimum magnetic field to be greater than $5 \times 10^{8}$ G. If the magnetic field is too low, the accretion disk could extend to the NS surface, for which we currently lack a robust solution. Therefore, the inner radius of the accretion disk may be slightly larger than the true value, leading to theoretical spin periods that are somewhat larger than the observational data.
 \begin{figure}
	\centering
	 \includegraphics[width=\linewidth]{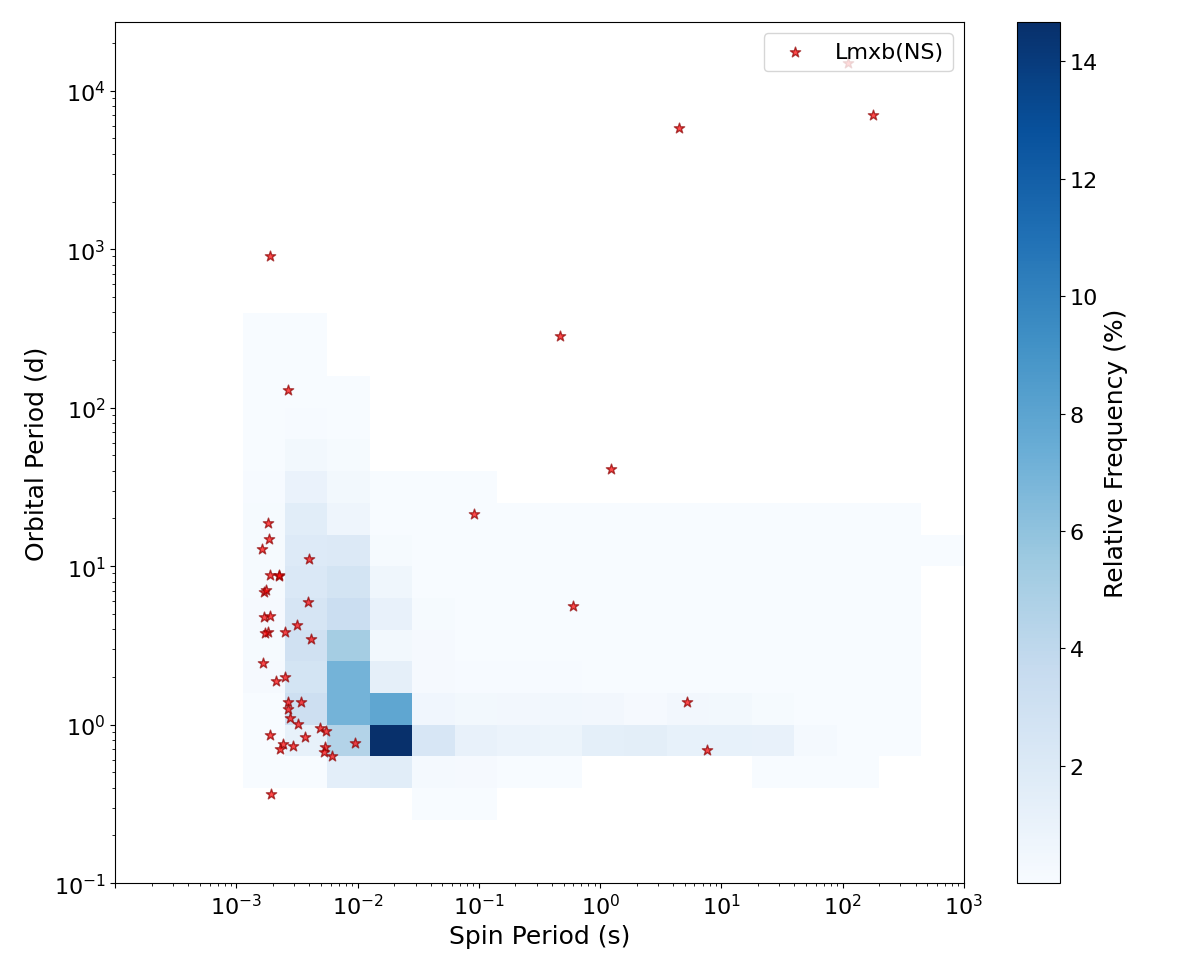}
	\caption{The distribution of NS spin periods vs. orbital periods for the low-mass X-ray binaries, where the color intensity indicates the probability of occurrence within each bin. The NS spin frequency and orbital period data corresponding to the pentagram symbol are taken from \citet{fortin2024catalogue}. 		
		\label{fig8.png}}
\end{figure}
\section{Summary}
\label{sect:analysis}
Using the BSE and MESA codes, along with the population synthesis method, we constructed a super-Eddington accretion NS model and investigated the mass growth of accreting neutron stars. In contrast to most previous studies, the neutron star mass in our research can effectively increase, even by up to $1.0 M_\odot$. The primary reason lies in the fundamentally different torques exerted on the neutron star's spin by the accretion disk under the two models, in our model, the neutron star remains in the accretion equilibrium phase for an extended period, which is highly conducive to mass growth.

In our model, the mass growth of neutron stars depends on the binary orbital period and the mass of the donor star. Our results can successfully account for the bimodal distribution of NS masses. The peak distribution of NS masses at around $\sim 1.8\,M_{\odot}$ primarily originates from NS binary systems where the donor star mass is less than $\sim 1.6\,M_{\odot}$ and the orbital period is shorter than 20 days. In these systems, the neutron star can remain in a quasi-equilibrium sub-Eddington accretion phase for an extended period. For NS binary systems with donor star masses between $\sim 2.0$ and $2.8\,M_{\odot}$ and orbital periods shorter than $\sim 1.6$ days, the accreting neutron star can achieve a mass exceeding $2.0\,M_{\odot}$. Meanwhile, NS systems that undergo common envelope evolution can account for the mass peak at $1.4\,M_{\odot}$, and these neutron stars can provide a channel for the mass peak at $1.4\,M_{\odot}$. In this study, we do not consider neutron stars formed via accretion-induced collapse (AIC). For NS binary systems originating from this channel, the companion star may not reside on the main sequence at the time of the NS's birth, and we have not quantified the fractional contribution of this population. Furthermore, the birth masses of neutron stars may not be strictly clustered around $1.4\,M_{\odot}$, but could instead follow a bimodal distribution .\citep{boccioli2024remnant}.

Moreover, many uncertainties remain in our model: the accretion efficiency $f_{\mathrm{mt}}$, the decay of the NS magnetic field, and the NS moment of inertia, among others. For instance, in our study, we assume the moment of inertia of the pulsar to be $10^{45}\,\mathrm{g}\,\mathrm{cm}^{2}$, which is almost the low limit for NSs. In reality, the actual moment of inertia of massive NSs may exceed $3 \times 10^{45} \ \mathrm{g \ cm^{2}}$ \citep{greif2020equation}. In terms of properties, for the model proposed by \citet{li2021maximum}, a larger I implies a slower spin-up process and a larger corotation radius compared to pulsar models with lower I values, which may allow the model to accrete more mass. For the model by \citet{shakura1973black}, a larger I may lead to a prolonged timescale for reaching equilibrium, and the pulsar might enter the long-term propeller regime earlier. These factors make the effect of variations in I on the accreted mass uncertain. On the other hand, due to uncertainties in the equation of state of pulsars, theoretical studies such as those by \citep{gao2022rotation} suggest that NSs with different equations of state exhibit different moments of inertia, which may influence the accretion process of pulsars. The study by \citet{zhong2025role} investigates the spin-up process of millisecond pulsars in both neutron star and strange star (SS) models. Moreover, during the accretion process of pulsars, a quark phase transition may occur, and the equation of state of NSs could evolve. Therefore, a more detailed study of accretion in pulsars with different equations of state is necessary.

\label{lastpage}

\end{document}